\documentclass[a4paper,11pt]{article}
\pdfoutput=1

\usepackage{amsmath}
\usepackage{amsfonts}
\usepackage{amssymb}
\usepackage{graphicx, rotating}
\usepackage{epstopdf}
\usepackage{epsfig}
\usepackage{latexsym}
\usepackage{multirow}
\usepackage{color}
\usepackage[dvipsnames]{xcolor}
\usepackage{slashed}
\usepackage[top=2.8 cm, bottom=2.8 cm, left=3 cm, right=3 cm]{geometry}
\usepackage{amstext} 
\usepackage{array} 
\usepackage[colorlinks]{hyperref}
\usepackage{tabularx}
\usepackage{authblk}
\usepackage{colortbl}
\usepackage{cite}

\usepackage{subcaption}

\usepackage{graphicx}
\usepackage{dcolumn}
\usepackage{bm}
\usepackage{hyperref}
\usepackage{color}
\usepackage{amsmath}
\usepackage{amssymb}
\usepackage{amsfonts}
\usepackage{multirow}

\usepackage{graphicx}
\usepackage{epstopdf}
\usepackage{dcolumn}
\usepackage{bm}
\usepackage{hyperref}
\usepackage{color}
\usepackage{amsmath}
\usepackage{scalerel}
\usepackage{url}
\usepackage{ulem}
\usepackage{lipsum}
\usepackage{mleftright}
\mleftright

\usepackage{lineno}

\newcommand{\vev}[1]{ \left\langle {#1} \right\rangle }

\def\tstau{{\tilde \tau}'}

\def\Mpl{M_{\rm pl}}

\hypersetup{citecolor={blue}}

\newcolumntype{L}{>{$}l<{$}} 
\newcommand{\ba}{\begin{eqnarray}}
\newcommand{\ea}{\end{eqnarray}}

\newcommand{\be}{\begin{equation}}
\newcommand{\ee}{\end{equation}}

\usepackage{hyperref}
\hypersetup{
colorlinks = true,
linkcolor = black,
citecolor={blue}
}
\usepackage{lineno}

\def\tstau{{\tilde \tau}'}

\begin{document}

\vspace{1cm}
\begin{titlepage}
\vspace*{-1.0truecm}
\begin{flushright}
CERN-TH-2022-020\\
 \vspace*{2mm}

 \end{flushright}
\vspace{0.8truecm}

\vspace{1mm}
\begin{center}
\huge\bf
Charged Dark Matter in \\ Supersymmetric Twin Higgs models
\end{center}

\vspace{1mm}
\begin{center}
\bf 
Marcin Badziak$^a$, Giovanni Grilli di Cortona$^{b}$,\\  Keisuke Harigaya$^{cd}$, and Micha\l\ \L ukawski$^{a}$\\

\vspace{5mm} 
{\small\sl $^a${\sl Institute of Theoretical Physics, Faculty of Physics, University of Warsaw, ul. Pasteura 5, PL-02-093 Warsaw, Poland  \vspace{0.2truecm}}

$^b${\sl Istituto Nazionale di Fisica Nucleare, Laboratori Nazionali di Frascati, C.P. 13, 00044 Frascati, Italy  \vspace{0.2truecm}}

$^c${\sl Theoretical Physics Department, CERN, Geneva, Switzerland \vspace{0.2truecm}}

$^d${\sl School of Natural Sciences, Institute for Advanced Study, Princeton, New Jersey 08540, USA \vspace{0.2truecm}}}
\end{center}

\begin{abstract}
\noindent Supersymmetric Twin Higgs models ameliorate the fine-tuning of the electroweak scale originating from the heavy scalar top partners required by the non-discovery of them at the Large Hadron Collider. If the Lightest Supersymmetric Particle resides in the twin sector, it may play the role of dark matter even if it is charged under twin gauge interactions. We show that the twin stau is a viable candidate for charged dark matter, even if the twin electromagnetic gauge symmetry is unbroken, with thermal relic abundance that naturally matches the observed dark matter abundance. A wide parameter space satisfies all the experimental constraints including those on dark matter self-interactions. Twin stau dark matter can be observed in future direct detection experiments such as LUX-ZEPLIN. The stau has a mass in the range of 300-500 GeV, and in the minimal scenario, has a decay length long enough to be observed as a disappearing track or a long-lived particle at the Large Hadron Collider.  
\end{abstract}

\end{titlepage}

\newpage

\begingroup
\color{black} 
\tableofcontents
\endgroup

\section{Introduction}
The electroweak (EW) scale is unnaturally small in comparison with other energy scales of the fundamental law such as the Planck scale. Such a hierarchy may be explained by supersymmetry (SUSY) with the masses of the superpartners of the Standard Model (SM) particles around the EW scale~\cite{Maiani:1979cx,Veltman:1980mj,Witten:1981nf,Kaul:1981wp}. No such particles have been found so far, and their masses should be above the TeV scale~\cite{ATLAS:2020dsf,CMS:2021beq}. Because of the large masses of superpartners, in the minimal embedding of the SM into a supersymmetric theory called the minimal supersymmetric SM (MSSM), the EW scale is obtained by fine-tuning of the parameters of the theory.

The Twin Higgs (TH) mechanism~\cite{Chacko:2005pe} can relax the required degree of fine-tuning in supersymmetric theories~\cite{Falkowski:2006qq,Chang:2006ra,Craig:2013fga,Badziak:2017syq,Badziak:2017kjk,Badziak:2017wxn} and provide a better understanding on the origin of the EW scale. The mechanism is based on a $\mathbb{Z}_2$ symmetry and predicts twin partners of the SM particles as well as those of the superpartners of the SM particles. As in the MSSM, the Lightest Supersymmetric Particle (LSP) is a dark matter (DM) candidate. The LSP may be the twin superpartner of a SM particle. Ref.~\cite{Badziak:2019zys} investigates the possibility that the twin bino-like LSP is DM and studies its phenomenology such as DM direct detection.

In this paper, we point out that the LSP may be a twin charged particle, such as a twin stau, a twin stop, and a twin gluino.  Although they are charged, they do not have a long-range force nor strong interactions with SM particles, and are good DM candidates. If the LSP is twin-colored, assuming that the abundance is determined by the freeze-out of its annihilation, its mass is required to be above a few TeV.%
\footnote{A lower mass is, however, possible if the LSP is non-thermally produced after the freeze-out.}
Since the Higgsino mass must be also above a few TeV in this case, the supersymmetric Higgs mass parameter is large and the small EW scale requires fine-tuning.  We thus focus on the twin stau LSP, whose mass may be just a few 100 GeV for a successful freeze-out scenario.

DM in the framework of Twin Higgs has been intensively studied in recent years. The role of DM may be played by twin neutrons, twin neutral atoms, or twin electrons~\cite{Farina:2015uea,Prilepina:2016rlq,Barbieri:2016zxn,Barbieri:2017opf,Chacko:2018vss,Koren:2020FreezeTwin,Chacko:2021vin,Beauchesne:2021opx}. Many more DM candidates have been proposed in various versions of Fraternal Twin Higgs model~\cite{Craig:2015pha}, e.g.,~twin taus, neutrinos, mesons, or bottom baryons~\cite{Garcia:2015THWimp,Garcia:2015toa,Craig:2015xla,Farina:2015uea,Freytsis:2016dgf,Prilepina:2016rlq,Cheng:2018vaj, Hochberg:2018vdo,Terning:2019hgj,Curtin:2021spx,Kilic:2021zqu}. In models of twin charged DM proposed so far, such as twin electrons or twin taus, the twin electromagnetic gauge symmetry is either broken or not present at all; otherwise long-range interactions mediated by twin electromagnetism would exclude these DM candidates. A novel feature of twin stau DM is its long-range self-interaction that is compatible with astrophysical data. This is possible due to soft supersymmetry breaking contributions to its mass $\gtrsim \mathcal{O}(100)$ GeV.

This paper is organized as follows. In Sec.~\ref{sec:TH}, we review supersymmetric Twin Higgs models and explain how the twin stau can be the LSP. In Sec.~\ref{sec:constraints}, experimental and observational constraints on the twin stau LSP are formulated. The lifetime of the MSSM stau, which is important for collider constraints and the computation of the thermal twin stau abundance, is computed in Sec.~\ref{sec:life}. In Sec.~\ref{sec:results}, we put together the constraints and show the viable parameter space and future prospects. The compatibility of the scenario with various solutions to the dark radiation problem in the Twin Higgs mechanism is discussed in Sec.~\ref{sec:DR}. In some case, the LSP mass may be correlated with the abundance of dark radiation. Finally, Sec.~\ref{sec:conclusion} is devoted to the conclusions.

\section{Supersymmetric Twin Higgs models}
\label{sec:TH}
 
In this section, we briefly review SUSY Twin Higgs models and discuss the possibility of the twin stau LSP.

\subsection{Higgs potential and tuning}

Two ingredients are needed in successful SUSY TH models with a natural EW scale:
\begin{itemize}
    \item 
    The potential of the SM and twin Higgses must be approximately SU(4) symmetric, so that the SM Higgs becomes a pseudo Nambu-Goldstone boson associated with the spontaneous breaking of the SU(4) symmetry by the twin Higgs, and as a result, the correction to the SM Higgs mass parameter is suppressed. The quadratic terms are automatically SU(4) symmetric because of the $\mathbb{Z}_2$ symmetry. The minimal quartic terms given by the D term of the EW and twin EW gauge interactions, on the other hand, are not SU(4) symmetric. Therefore, an additional large SU(4) invariant quartic term is necessary.
    \item
    The SM Higgs mass of 125 GeV must be obtained without large stop masses that lead to large tuning of the Higgs mass parameter.
\end{itemize}
In the original SUSY Twin Higgs model~\cite{Falkowski:2006qq,Chang:2006ra}, an F-term of a singlet chiral field $S$ generates an SU(4) invariant quartic term via a superpotential $W \sim S(H_u H_d + H_u' H_d')$. The EW D-term generates an SU(4)-breaking quartic coupling which contributes to the Higgs mass. These quartic terms depend on the ratio of the vacuum expectation value (VEVs) of the up-type Higgs to that of the down-type Higgs, $\tan\beta$  (assumed to be the same for the twin sector). If one takes a large $\tan\beta$ to increase the tree-level Higgs mass from the D-term, so that the required stop mass is small, the $SU(4)$-invariant F-term becomes small. In this case, the Twin Higgs mechanism does not work well. On the other hand, taking small $\tan\beta$ allows for a large $SU(4)$ symmetric quartic term, at the price of a small tree-level Higgs mass. This requires a large stop mass. As a result, the EW scale is fine-tuned at least at the level of 1\%~\cite{Craig:2013fga}.

An alternative was proposed in~\cite{Badziak:2017syq}, leading to fine-tuning better than a few percent and, in some of the parameter space, as good as ten percents. The MSSM and its twin copy are supplemented by an abelian U(1)$_X$ gauge symmetry generating an $SU(4)$-invariant D term. The U(1)$_X$ is spontaneously broken, but if the masses of the symmetry breaking fields are dominated by the SUSY breaking soft mass, the D-term potential does not decouple. In this model the quartic coupling is maximised in the limit of large $\tan\beta$, and the Higgs mass of 125 GeV is easily obtained without large stop masses. In order to obtain a large SU(4) invariant quartic, the U(1)$_X$ gauge coupling is required to be large and consequently its Landau pole appears below the Planck scale. However, the U(1)$_X$ gauge symmetry may be embedded into a non-abelian group, so that the Landau pole is above the Planck scale or the theory is asymptotically free~\cite{Badziak:2017kjk,Badziak:2017wxn}. The model is then fully perturbative below the Planck scale.

\subsection{The lighest supersymmetric particle}

If R-parity is conserved, the LSP is absolutely stable.
In principle, the LSP may be a superpartner of a twin particle rather than that of a SM particle. The twin bino LSP has been already investigated in~\cite{Badziak:2019zys}. In this paper, we consider the twin stau LSP.

Assuming that soft SUSY breaking terms in the MSSM and the twin sector are the same due to the $\mathbb{Z}_2$ symmetry, the physical masses of the MSSM and the corresponding twin states differ only due to the larger vacuum expectation value (VEV) of the twin Higgs, $v'$, compared to that of the SM Higgs, $v$. For $\tan\beta>1$, both the right-handed and left-handed (twin) stau obtain a positive mass squared from the D-term coupling with the (twin) Higgs. As a consequence, the pure right-handed and left-handed twin staus are heavier than their MSSM counterparts. Moreover, for the pure left-handed twin stau the twin sneutrino is even lighter and play the role of the LSP.

Nevertheless, the twin stau can be the LSP because of the mixing between the left-handed and right-handed states. The off-diagonal entry of the squared mass matrix of MSSM (twin) staus is proportional to $m_{\tau (\tau')}\left(A_\tau-\mu\tan\beta\right)$, where $\mu$ is the Higgsino mass parameter and $A_\tau$ the soft SUSY breaking tau trilinear coupling, taken to be the same for the MSSM and twin sector. Hence, the mixing effects are bigger in the twin stau sector, favouring a twin stau LSP. The twin stau LSP is either mostly right-handed or a highly-mixed left-right state. It can never be strongly dominated by the left-handed component because in such a case twin sneutrino would be the LSP.

\section{Experimental constraints}
\label{sec:constraints}
In this section, we discuss the experimental and observational constraints on the twin stau LSP. The resulting constraints on the parameter space of the theory are presented in Sec.~\ref{sec:results}.

\subsection{Thermal abundance}

We will focus on a mass spectrum such that the LSP is the twin stau and the twin Higgsino is not too heavy in order to obtain the EW scale without fine-tuning. We also allow the bino to be light, since a light bino has important phenomenological implications. The wino can also affect the relic density but its effect is minor in comparison with the bino except for the parameter space where the LSP has an $\mathcal{O}(1)$ left-handed twin stau component. We thus mainly take a large wino mass for simplicity, but comment on the impact of a light wino. All the other twin and MSSM particles (squarks, first and second generation sleptons, and gluino) do not affect the relic density computation as long as they are not degenerate with the stau so that coannihilation with them is ineffective. We mostly consider the case where the twin and the MSSM tau Yukawa couplings are equal, but also comment on the case with different couplings.

The dominant twin stau annihilation channels resemble those of the MSSM stau~\cite{Ellis:1998kh,Ellis:1999mm,Pradler:2008qc} with the MSSM states replaced by the corresponding twin states. The twin stau annihilation channels depend mainly on the size of the left-right mixing. For the right-handed dominated twin stau the dominant annihilation channel is into twin photons unless the twin bino is light in which case annihilation into twin taus may become dominant. Finally, in the presence of a larger left-handed twin stau component of the LSP, the annihilation into twin $W$'s becomes relevant.

We compute the relic density with our modified version of \texttt{Micromegas}~\cite{Belanger:2001fz,micromegasMSSM,Belanger:2006is}. In our setup \texttt{Micromegas} automatically takes into account the coannihilation within particles in the twin sector. Coannihilation with the MSSM sector becomes relevant for the relic density computation if there are states with masses relatively close to the twin stau, $\Delta_i = (m_i - m_{\tstau})/m_{\tstau} \lesssim 0.05$. In the simplified case in which all superpartners except for MSSM and twin staus are decoupled, only the MSSM lightest stau and the twin sneutrino may satisfy this condition. 

In order to capture the effect of the coannihilation with the MSSM stau, we use the approximation that the annihilation cross section $\sigma$ of the twin stau and the MSSM stau are the same, which is justified by the approximate $\mathbb{Z}_2$ symmetry.
We also neglect mixed annihilation cross-section between the MSSM stau and the twin stau.%
\footnote{
Such annihilation can be induced via the Higgs portal, Higgsino and bino mixing discussed in Sec.~\ref{sec:life}, or the $X$ gaugino portal,  but is suppressed by a loop factor, smallness of the mixing, or the heavy $X$ gaugino mass, respectively. The resultant cross-section is hence much smaller than the twin stau annihilation cross-section that is induced at the tree-level with $\mathcal{O}(1)$ couplings without large mass scales.
}

We then compute the effective annihilation cross section following~\cite{Griest:1990kh}, which is given by
\begin{equation}
 \sigma_{\mathrm{eff}} =\sigma \frac{1+(1+\Delta)^3 e^{-2 x_f \Delta}}{[1+(1+\Delta)^{3/2} e^{-x_f \Delta}]^2},
\end{equation}
where $x_f=m_{\tstau}/T_f\sim25$, with $T_f$ the freeze-out temperature. Assuming that the effects of coannihilation on the freeze-out temperature are subdominant, the relic density can be simply computed by the scaling of the abundance without the coannihilation $\Omega h^2$ by the cross sections, $\Omega_{\mathrm{coann}} h^2 = \Omega h^2 \sigma/\sigma_{\mathrm{eff}}$. Within this simplified scheme, the effects of the coannihilation can be controlled by the masses of the spectrum alone.

The most interesting case is when the thermal relic abundance of the twin stau LSP computed above, $\Omega^{\rm th} h^2$,  fits the observed density of dark matter $\Omega_{\rm obs} h^2\approx0.12$ \cite{Planck:2018vyg}. We refer to this case as thermal DM. Nevertheless, in analysing the constraint from and sensitivity of DM direct detection, we also consider two other possible scenarios in which the abundance of the twin stau LSP, $\Omega_{\tstau}$, may be different from its thermal abundance:
\begin{itemize}
    \item Non-thermal DM with $\Omega_{\tstau}=\Omega_{\rm obs}$, which may be achieved in a non-standard cosmological history with non-thermal DM production (for $\Omega^{\rm th}<\Omega_{\rm obs}$) or entropy production (for $\Omega^{\rm th}>\Omega_{\rm obs}$).
    \item Multi-component DM with $\Omega_{\tstau} = \Omega^{\rm th} <\Omega_{\rm obs}$. In this scenario it is assumed that apart from the twin stau LSP there exist an additional component of DM which ensures that the total relic density fits the observed value.
\end{itemize}
See~\cite{Harigaya:2014waa} for the up-to-date estimation of non-thermal DM abundance with the finiteness of the thermalization time scale properly taken into account. The non-standard cosmological history can solve the dark radiation problem in the Twin Higgs mechanism as we discuss in Sec.~\ref{sec:DR}.

\subsection{Twin stau self-interactions} 

The twin stau possesses a twin electromagnetic charge, and hence DM in our scenario has long-range self-interactions. This has interesting implications on galaxy formation. In particular, the strongest constraint comes from the triaxial structure of galaxy halos. The self-interactions can create a more isotropic dark matter velocity distribution in two ways. On one hand, interactions with large momentum exchange may immediately reduce anisotropy significantly. On the other hand, the cumulative effect of many soft interactions makes the velocity distribution more isotropic. The resulting constraints have been analysed in Ref.~\cite{Agrawal:2016quu} for a Dirac fermion charged under a dark U(1) gauge symmetry, using measurements of the non-zero ellipticity of the gravitational potential of NGC720. Those results can be straightforwardly applied to the case of twin stau DM since the self-scattering cross-section of the twin stau in the non-relativistic limit  is the same as that for a Dirac fermion. Assuming that the twin and the SM electromagnetic gauge couplings are equal, the ellipticity constraint sets a lower bound on the twin stau mass of about 210~GeV. As we will see, the twin stau mass leading to the correct thermal abundance of DM is above this bound. Notice, however, that this bound should be taken with caution. \\

Previously, the same data have been analysed in Ref.~\cite{Feng:2009mn} resulting in a much stronger lower bound on the twin stau mass of about 1~TeV. However, Ref.~\cite{Agrawal:2016quu} has made several improvements in extracting the constraint with respect to Ref.~\cite{Feng:2009mn}. In particular, Ref.~\cite{Feng:2009mn} neglects the saturation effect in the evolution of the ellipticity, approximating the change of the ellipticity linearly. 
Moreover, the constraint should be taken at larger radius with lower DM density, further suppressing the bound. A detailed analysis of the differences between these constraints can be found in \cite{Agrawal:2016quu}. We will use the more recent bound, that however should be taken with caution. Indeed, there are many challenges to the ellipticity measurements on DM self-interactions.

For smaller dark matter abundance, the characteristic timescale for a particle to change its kinetic energy by an $\mathcal{O}(1)$ factor, interpreted as the timescale on which the velocity completely randomizes, becomes longer. Therefore, for $\Omega_{\tstau} h^2 \lesssim 0.12$, the lower bound on the twin stau mass from self-interactions is even below $210$~GeV.

\subsection{Direct detection}

A significant portion of the parameter space of the model can be probed via direct-detection experiments.
The interaction of the twin stau DM with nucleons is mediated by the Higgs portal and kinetic mixing.%
\footnote{In D term SUSY TH models, $Z$ and $Z'$ mix with each other via $X$, and the resultant $Z$-$Z'$ mixing also contributes to direct detection. This contribution can be comparable to the Higgs portal if the X boson is sufficiently light.}

\subsubsection{Higgs portal}

The twin stau couples to the twin Higgs, which mixes with the SM Higgs. The effective Higgs-twin stau coupling reads
\begin{eqnarray}
\lambda_{h\tilde{\tau}'\tilde{\tau}'} &=&\frac{g}{m_{W'}}\left[ \left(-\frac{1}{2} c_{\theta_{\tilde{\tau}'}}^2+s_W^2 c_{2\theta_{\tilde{\tau}'}}\right) m_{Z'}^2 s_{\alpha+\beta} + m_{\tau'}^2 \frac{s_\alpha}{c_\beta} + \frac{m_{\tau'}}{2}\left( A_{\tau}\frac{s_\alpha}{c_\alpha} + \mu \frac{c_\alpha}{c_\beta} \right)s_{2\theta_{\tilde{\tau}'}}  \right]   \frac{v}{v'} \nonumber\\
&\simeq& \frac{g}{m_{W'}}\left[ \left(\frac{1}{2} c_{\theta_{\tilde{\tau}'}}^2-s_W^2 c_{2\theta_{\tilde{\tau}'}}\right) m_{Z'}^2 c_{2\beta} - m_{\tau'}^2 - \frac{m_{\tau'}}{2}\left( A_{\tau} - \mu \tan\beta \right)s_{2\theta_{\tilde{\tau}'}}  \right]   \frac{v}{v'}
\label{eq:hstau_coup},
\end{eqnarray}
where $s_{x}$ and $c_x$ stands for $\sin x$ and $\cos x$ respectively, $\tan\beta=v_u/v_d$ is the ratio between the VEVs acquired by $H_u$ and $H_d$, and $\alpha$ is the mixing angle necessary to diagonalize the neutral CP-even Higgs mass matrix. The masses of the twin tau, W, and Z are denoted by $m_{\tau'}$, $m_{W'}$ and $m_{Z'}$ respectively, while $g$ is the electroweak coupling constant. The second line is obtained in the decoupling limit by using $\cos(\beta-\alpha)=0$. The twin stau mixing angle $\theta_{\tilde{\tau}'}$ diagonalizes the twin stau mass matrix and therefore relates the mass eigenstates $\tilde{\tau}_1'$ and $\tilde{\tau}_2'$  to $\tilde{\tau}_L'$ and $\tilde{\tau}_R'$ via an orthogonal transformation
\begin{equation}
    \begin{pmatrix}
      \tilde{\tau}_1' \\ 
      \tilde{\tau}_2'
    \end{pmatrix}
    = R_{\tilde{\tau}'}
    \begin{pmatrix}
      \tilde{\tau}_L' \\
      \tilde{\tau}_R'
    \end{pmatrix}
    \qquad\textrm{with}\quad
      R_{\tilde{\tau}'} =
    \begin{pmatrix}
      \cos{\theta_{\tilde{\tau}'}} & \sin{\theta_{\tilde{\tau}'}} \\
      -\sin{\theta_{\tilde{\tau}'}} & \cos{\theta_{\tilde{\tau}'}}
    \end{pmatrix}.
    \label{eq:tstau_mixing}
\end{equation}
 The twin stau thus interacts with nucleons via Higgs exchange, with a cross section
\begin{align}
\sigma_n = \frac{\mu_{\tilde{\tau}'N}^2 c_N^2}{4 \pi m_{\tilde{\tau}'}^2}, \,\,  \frac{c_N}{m_N} = \sum_q  \frac{c_q f_{T_q}^{(N)}}{m_q} + \frac{2f_{T_G}^{(N)}}{27}\sum_Q \frac{c_Q}{m_Q},
\end{align}
where $\mu_{\tilde{\tau}'N}$ is the nucleon-twin stau reduced mass, $m_{\tilde{\tau}'}$ is the twin stau mass, and $c_N$ is the twin stau-nucleon effective coupling. The sums run over $q=u,d,s$ and $Q=c,b,t$ and the coefficients $c_{q,Q}$ are given by
\begin{equation}
c_{u_i}= -\frac{\lambda_{h\tilde{\tau}'\tilde{\tau}'}}{m_h^2}\frac{m_{u_i}}{v}\frac{\sin\alpha}{\cos\beta},\qquad c_{d_i}= \frac{\lambda_{h\tilde{\tau}'\tilde{\tau}'}}{m_h^2}\frac{m_{d_i}}{v}\frac{\cos\alpha}{\sin\beta},
\end{equation}
for up- and down-type quarks, respectively.

\subsubsection{Kinetic mixing}
\label{subsubsec:kin_mix}
The twin photon may have a kinetic mixing with the SM photon,
\begin{align}
{\cal L} = - \frac{\epsilon}{2}F_{\mu \nu} F^{'\mu \nu}.
\end{align}
The twin stau interacts with nucleons via the kinetic mixing, and XENON1T~\cite{Aprile:2018dbl} gives a constraint~\cite{Dunsky:2019api}
\begin{align}
\label{eq:epsilon_DD}
\epsilon < 10^{-10} \left( \frac{m_{\tilde{\tau}'}}{100\,{\rm GeV}} \right)^{1/2}.
\end{align}
In the next section, we will consider a case where the bino and the twin bino have a mass mixing. For such a case, the strong upper bound on the kinetic mixing requires a structure of the theory such that the mass mixing is sizable while the kinetic mixing is suppressed.

\subsection{Collider}

\begin{figure}[t!]
\centering
\includegraphics[scale=0.25]{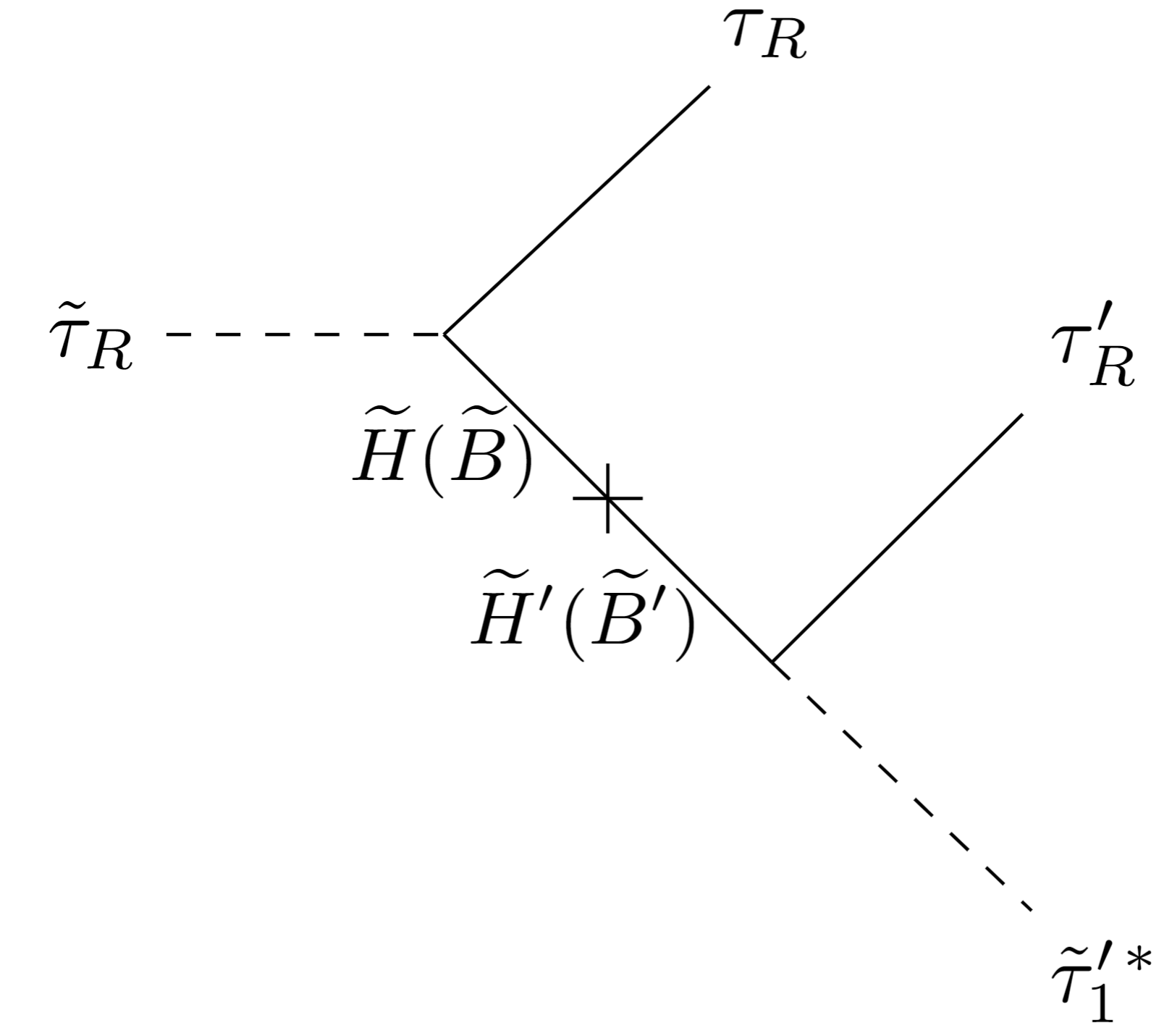}\hfill
\caption{The lightest stau decay is mediated by mixing between higgsino and twin higgsino and/or mixing between bino and twin bino which depends on the UV completion of the model.}
\label{fig:tstauRdecays}
\end{figure}

\begin{figure}[t!]
\begin{center}
 \begin{subfigure}{0.3\textwidth}
    \centering
\includegraphics[width=1\textwidth]{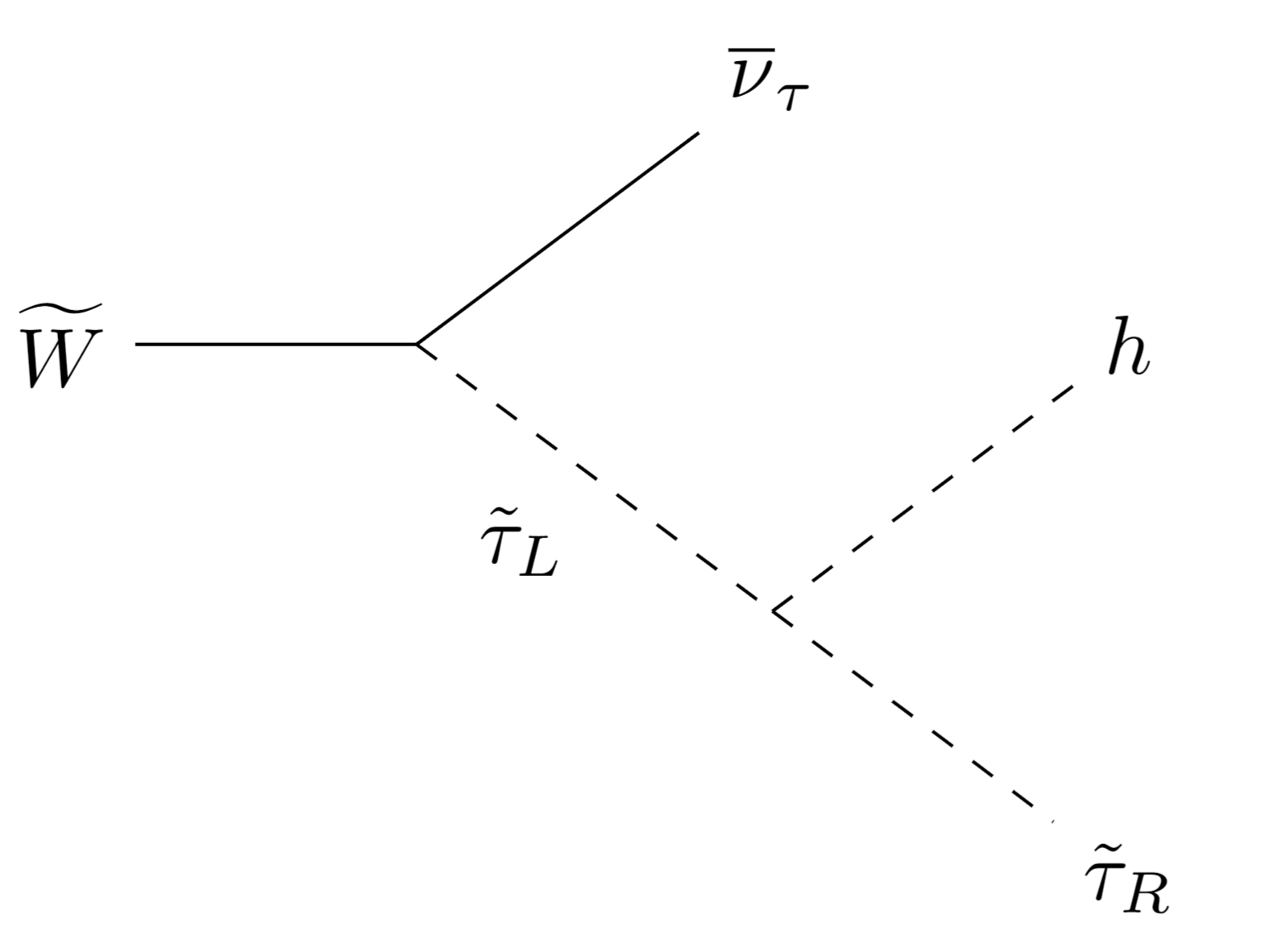}\hfill
    \caption{}
    \label{fig:winoDecays1}
\end{subfigure}
 \begin{subfigure}{0.3\textwidth}
    \centering
\includegraphics[width=1\textwidth]{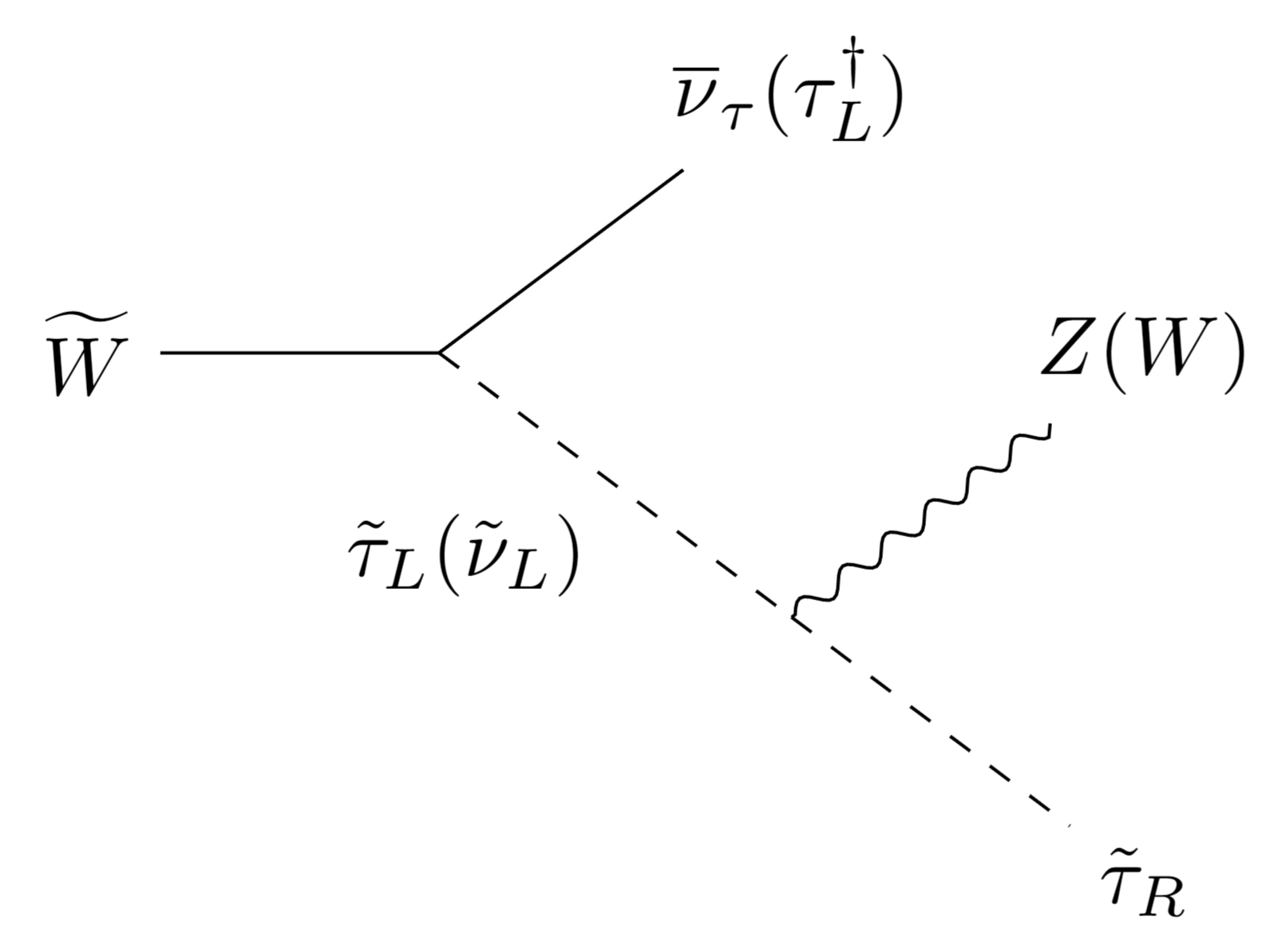}\hfill
    \caption{}
    \label{fig:winoDecays2}
\end{subfigure}
 \begin{subfigure}{0.3\textwidth}
    \centering
\includegraphics[width=1\textwidth]{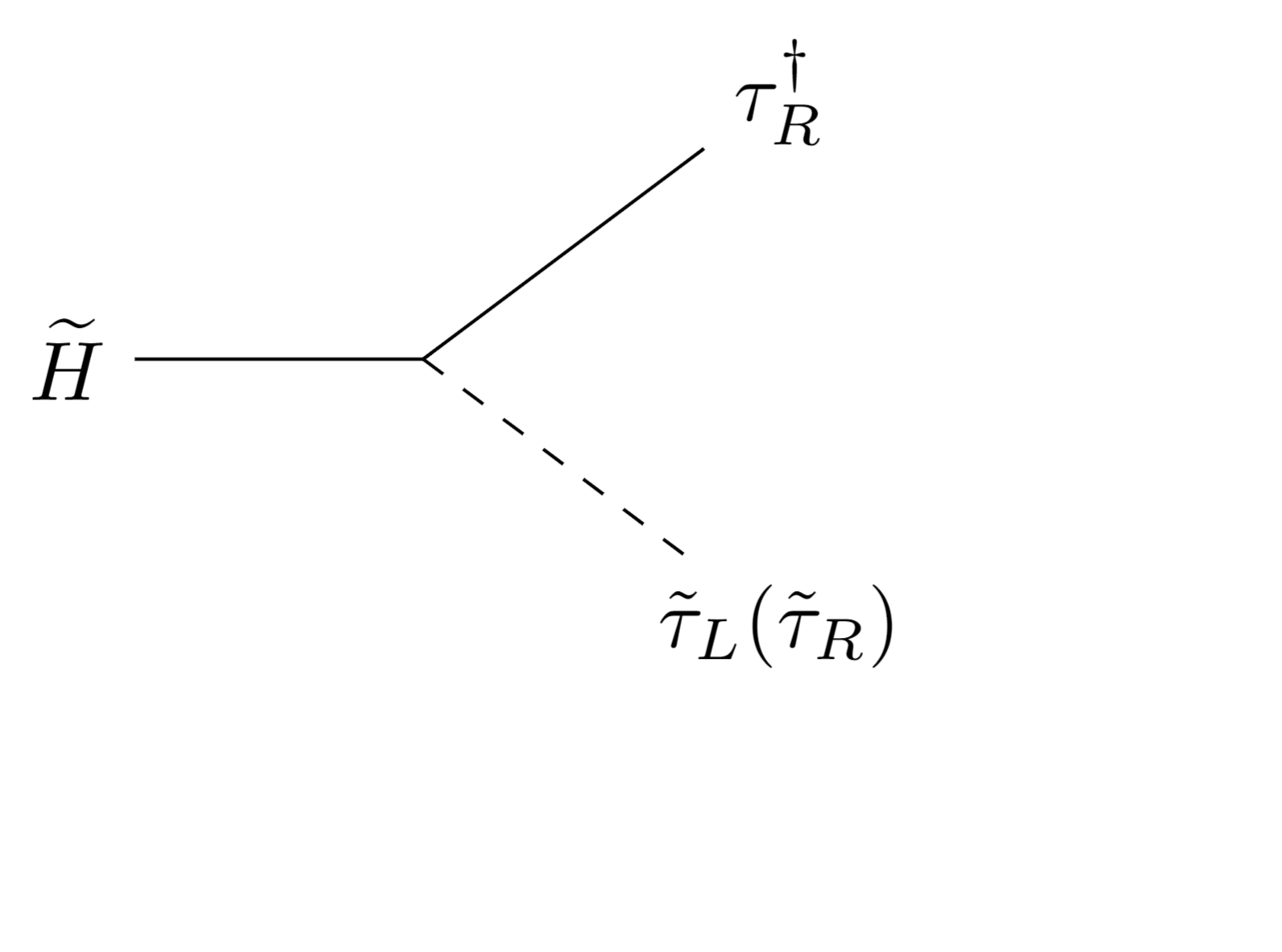}\hfill
    \caption{}
    \label{fig:higgsinoDecay}
\end{subfigure}
\caption{Main wino and higgsino decay channels. All decay channels of MSSM particles end in a state involving the lightest stau which decays into tau, twin tau and twin stau, as shown in Fig. \ref{fig:tstauRdecays}.}
\end{center}
\end{figure}
The main collider signature of this scenario is the existence of relatively light staus in the range of several hundred GeV. The experimental constraints on the stau mass significantly depends on the lifetime of the stau. LHC constraints are absent for promptly decaying staus, because the mass splitting between the stau and the twin stau LSP is small (few tens of GeV at most), resulting in soft taus in the final states. Moreover, the lightest stau in this scenario is dominated by the right-handed component and the LHC is not yet sensitive to direct production of right-handed staus even in the limit of massless LSP~\cite{CMS:2021woq}. We should also note that in the present scenario the decays of the lightest stau are three-body, as shown in Fig.~\ref{fig:tstauRdecays}, in contrast to the MSSM-motivated LHC stau searches assuming two-body decays of the stau, so the taus in the final states are even softer.

LHC constraints on the wino and Higgsino that decay into staus may also be relevant, but they are not constraining as long as staus decay promptly. The wino decay proceeds as follows: the wino decays into a tau/neutrino and a left-handed slepton, which subsequently decays into $Z/W/h$ and a right-handed stau, as shown in Figs.~\ref{fig:winoDecays1} and \ref{fig:winoDecays2}, with the right-handed stau decaying into a tau, a twin tau, and a twin stau, see Fig.~\ref{fig:tstauRdecays}.  The search that exploits taus and missing energy does not constrain the parameter region where the LSP mass is above 100~GeV~\cite{CMS:2021cox}, which is the region of parameter space relevant for twin stau DM. The search using $Z/W/h$ and missing energy could give a stronger constraint, but the lower bound on the wino mass is 600 GeV even with a massless LSP, and the constraint is absent if the LSP mass is above 300 GeV~\cite{CMS:2021cox,ATLAS:2021moa}. In our setup, the left-handed slepton is boosted, so the $Z/W/h$ tends to be aligned with the tau from the right-handed stau decay, and leptons from $Z/W/h$ may be vetoed by the isolation criteria imposed in the search, weakening the bound. If the first two generation sleptons are also lighter than the wino, the lower bound on the wino mass becomes stronger and is 1300 GeV in the limit of massless LSP. This limit, however, is absent for the LSP mass above 800~GeV~\cite{CMS:2021cox}.

The Higgsino decays into a tau and a left-handed slepton, or a tau and a right-handed slepton, see Fig.~\ref{fig:higgsinoDecay}. Both give the same signal as the wino discussed above, except that the decay into a tau and a right-handed lepton does not involve $Z/W/h$. Because of the smaller cross section, the constraint on the Higgsino mass is even weaker. Even with the first two generation of sleptons lighter than the third generation, the bound on the Higgsino mass is expected to be weak since the Higgsino dominantly decays into a tau and a stau.

The left-handed stau is required to be light so that the twin stau becomes the LSP via left-right mixing. The left-handed stau decays into $Z/W/h$ and a right-handed tau, which subsequently decays into a tau, a twin tau, and a twin stau. The topology of the event is similar to that of the wino decay, but the constraint on the left-handed stau mass is absent because of the smaller cross section.

Finally, for sufficiently long-lived staus the LHC constraints on the stau mass and the masses of other MSSM particles that decay into stau become more important. Since the lifetime of the stau affects not only the collider phenomenology of the model but also the relic density computation, we discuss it in some more detail in the next section.

\section{The lifetime of stau}
\label{sec:life}

In this section, we discuss the lifetime of the stau, which affects the phenomenology in several ways.
\begin{itemize}
\item
The stau may be observed as a long-lived charged particle at collider experiments. The LHC searches for long-lived charged particles such as stau are the most sensitive if the lifetime is longer than about 10 ns (that corresponds to the decay length of $\mathcal{O}(1)$~m)~\cite{Evans:2016zau}. In such a case the lower bound on the stau mass is about 430 GeV for the right-handed stau~\cite{Aaboud:2019trc}. Most of the parameter region with successful thermal freeze-out is then excluded, as we will show in section \ref{sec:results}. Furthermore, since the stau is also produced from the decay of heavier MSSM particles, their mass is also bounded from below. For example, the stop, gluino, wino, Higgsino, and left-handed slepton masses must be above about 1600, 2200, 1300, 1200, and 500 GeV, respectively.

For the stau decay length between about 10~cm and 1~m LHC searches for disappearing charged tracks may become sensitive~\cite{CMS:2020atg,ATLAS:2021ttq}. The upper limit on the production cross-section at the LHC with $\sqrt{s}=13$~TeV can be as small as about 10~fb for the decay length $\mathcal{O}(1)$~m \cite{CMS:2020atg} which corresponds to the production cross-section of the 200 GeV right-handed staus.%
\footnote{The production cross-sections for staus can be found on the webpage of the SUSY cross-section Working group: https://twiki.cern.ch/twiki/bin/view/LHCPhysics/SUSYCrossSections}
Thus, the disappearing track searches are not yet sensitive to our scenario. To sum up, the LHC constraints are relevant only if the stau decay length is above $\mathcal{O}(1)$~m and put lower bound on the masses of MSSM particles.

\item
The decay product of the stau contains a tau. If the lifetime is longer than $10^5$~s, the abundance of the stau at the time of decay must be much smaller than the dark matter abundance, so that Big-Bang Nucleosynthesis (BBN) is not disturbed~\cite{Kawasaki:2017bqm}. Since the annihilation rate of the stau is comparable to that of the twin stau, the stau abundance before it decays is of the same order as the twin stau abundance. As a consequence the twin stau cannot be DM. We found, however, that the lifetime of stau is generically much shorter than $10^5$~s and BBN does not constrain our scenario, unless the stau is nearly degenerate with the twin stau.
\item
If the decay of the stau occurs after the freeze-out of the twin stau annihilation around $T\simeq m_{\tilde{\tau}'}/25$,   the abundance of the twin stau is enhanced. Since the stau abundance before it decays is of the same order as the twin stau freeze-out abundance, the final dark matter abundance is larger  by a factor of about 2 with respect to the freeze-out value. However, the coannihilation between stau and twin stau is no longer important if the stau decay is ineffective. This occurs if the lifetime is longer than
\begin{align}
    0.2 \frac{\Mpl}{(m_{\tilde{\tau}}/25)^2} \simeq \mathcal{O}(1)~{\rm ns} \left( \frac{300~{\rm GeV}}{m_{\tilde{\tau}}} \right)^2 \simeq \mathcal{O}(1)\, \mathrm{m} \left( \frac{300~{\rm GeV}}{m_{\tilde{\tau}}} \right)^2.
\end{align}
\end{itemize}
We first discuss the case in which the stau lifetime is determined by the Higgsino-twin Higgsino mass mixing and significantly constrains the parameter space of the model. We then discuss possible bino-twin bino mass mixing and show that the lifetime can be short enough that none of the above issues arises. In both cases, the mediators of the decay include the bino. We assume that the lighter stau is dominantly the right-handed one. When the lighter stau contains a  significant  fraction  of  the  left-handed  twin  stau,  the  decay  rate becomes smaller by a factor of up to 16. The suppression, however, can be compensated by the change of parameters by $\mathcal{O}(1)$ factors. 

\subsection{Higgsino mass mixing}
Since the Higgs and the twin Higgs couple with each other via a quartic term, we expect that the Higgsino also couples with the twin Higgsino. As a result the neutral Higgsino has a mass mixing with the neutral twin Higgsino. The magnitude of the mixing depends on the UV model. In order to be concrete, we consider a D-term Twin Higgs model where the mixing is given by
\begin{align}
\label{eq:higgsino_mixing}
- \epsilon_{\tilde{H}} \left(\tilde{H}_u - \frac{1}{{\rm tan}\beta} \tilde{H}_d\right ) \left(\tilde{H}'_u - \frac{1}{{\rm tan}\beta} \tilde{H}'_d\right ).
\end{align}
Here the mixing $\epsilon_{\tilde{H}}$ is generated by the exchange of a heavy gaugino and is as small as $ v v' m_{\rm soft} / v_S^2$, where $v_S$ is the symmetry breaking scale of the gauge symmetry whose $D$ term is responsible for the $SU(4)$ invariant Higgs quartic potential and $m_{\rm soft}$ is a soft SUSY breaking mass of the symmetry breaking field. For natural choices of parameters we found that typically $\epsilon_{\tilde{H}}\sim\mathcal{O}(0.01-0.1)v'$.  A detailed computation is  given in Appendix~\ref{sec:App_higgsinomix}.

The main stau decay channel is $\tilde{\tau}\to \tilde{\tau}^{'\dag}\tau \tau'$. (Another decay channel $\tilde{\tau}\to \tilde{\tau}'\tau \tau^{'\dag}$ is p-wave suppressed.) In the limit of large $\tan\beta$ and $\mu,M_1 \gg g'v'$, we find the decay length
\begin{align}
\label{eq:decay}
\Gamma^{-1} \simeq& \left(\frac{1}{480\pi^3} \frac{(m_{\tilde{\tau}}-m_{\tilde{\tau}'})^5}{ M^2 m_{\tilde{\tau}}^2}\right)^{-1}\simeq 9\,{\rm ns} \left(\frac{m_{\tilde{\tau}}}{300~{\rm GeV}}\right)^2  \left(\frac{M}{10^6~{\rm GeV}}\right)^2 \left(\frac{10~{\rm GeV}}{m_{\tilde{\tau}}-m_{\tilde{\tau}'}}\right)^5, \nonumber \\
\frac{1}{M} =& \frac{g'^4 v v' \varepsilon_{\tilde{H}} m_{\tilde{\tau}}^2 (M_1^2 + m_{\tilde{\tau}}^2)}{(M_1^2-m_{\tilde{\tau}}^2 )^2  (\mu^2- m_{\tilde{\tau}}^2)^2},
\end{align}
where $M_1$ is the (twin) bino mass parameter.
Here we neglect the tau and twin tau masses, but we include them in our numerical results. The decay rate is suppressed for large $|M_1|$ and $|\mu|$. In order for the decay length of the stau to be below $1$ m via the Higgsino mixing, $M_1$ and $\mu$ must not be much above the stau mass.

The requirement that the stau decays before freeze-out can be achieved with an interplay between the different parameters involved in the decay rate: $M_1$, $\mu$, $\tan\beta$,  $m_{3L}$, $m_{3R}$, and $A_\tau$. The parameter $M_1$ can be as small as the stau mass. However, this unavoidably leads to twin stau annihilation via twin bino exchange which leads to a larger mass for the twin stau LSP consistent with $\Omega h^2\approx0.12$. The parameters $\mu$, $\tan\beta$,  $m_{3L}$, $m_{3R}$ and $A_\tau$ are all related and affect the decay width with competing effects. The $\mu$ parameter is bounded to be below $\sim 1.5$ TeV by the need to solve the naturalness problem. In principle, small $\mu$ implies an enhanced decay rate. However, the smaller $\mu$ the smaller is the mixing between the left- and right-handed twin staus. This makes the twin stau heavier than the MSSM counterpart (if mostly right handed) or than the twin sneutrino (if mostly left handed) and the LSP is not the twin stau. This can be compensated by requiring large $\tan\beta$ or a $\mathbb{Z}_2$ breaking in the tau Yukawa couplings. We will show the allowed parameter space in section~\ref{sec:results}.

\subsection{Bino mass mixing}
The bino and twin-bino may have mass mixing,
\begin{align}
{\cal L} \supset  - \epsilon_M M_1\tilde{B}\tilde{B}' + {\rm h.c.}
\end{align}
A stau can then decay into a tau, a twin stau, and a twin tau via bino exchange. The effective operator describing the decay is
\begin{align}
\frac{1}{M}\bar{\tau} \bar{\tau}' \tilde{\tau} \tilde{\tau}',~~\frac{1}{M} = \frac{2g^{'2}  \epsilon_M}{M_1},
\end{align}
where$\bar{\tau}$ is the right-handed tau and we use the approximation $M_1 \gg m_{\tilde{\tau}} $. The lifetime of the stau is
\begin{align}
\label{eq:Gamma_binomix}
     \Gamma^{-1} = 2\,{\rm ns}\, \left(\frac{m_{\tilde{\tau}}}{300~{\rm GeV}}\right)^2  \left(\frac{M_1}{1000~{\rm GeV}}\right)^2 \left(\frac{10~{\rm GeV}}{m_{\tilde{\tau}}-m_{\tilde{\tau}'}}\right)^5 \left(\frac{10^{-2}}{\epsilon_M}\right)^2
\end{align}
and may be easily shorter than $10$ ns, so that the stau decays before the freeze-out and the collider constraints from long-lived particle searches are evaded.

While the mass mixing parameter $\epsilon_M$ must be large enough to make the stau short-lived enough, the photon-twin photon kinetic mixing must be suppressed to avoid too large a twin stau-nucleon scattering cross section, cfr section~\ref{subsubsec:kin_mix}. This can be achieved by a supersymmetry breaking field $Z$ with a coupling and VEV,
\begin{align}
\int {\rm d}\theta^2 \frac{Z}{M_*} W W',~~
\vev{Z} \ll M_*,~~ \vev{F_Z} = \epsilon_M M_1 M_*,
\end{align}
where $M_*$ is a cut-off scale. Although $\vev{Z}\sim M_*$ may be avoided by a discrete symmetry of $Z$,
non-zero VEV of $Z$ may be induced by supergravity effect and quantum corrections. We investigate those effect for an explicit example in Appendix~\ref{sec:App_binomix} and find that the VEV is sufficiently small and the kinetic mixing is much below the constraint from DM detection experiments shown in Eq.~(\ref{eq:epsilon_DD}).

\section{Results}\label{sec:results}

In this section we assemble together all the bounds described above and show representative plots. We start by considering the case of a decoupled (twin) bino and wino, with a Higgsino at $\sim 1.5$ TeV and $\tan\beta=10$. 
\begin{figure}[t]
\begin{center}
 \includegraphics[width=0.49\textwidth]{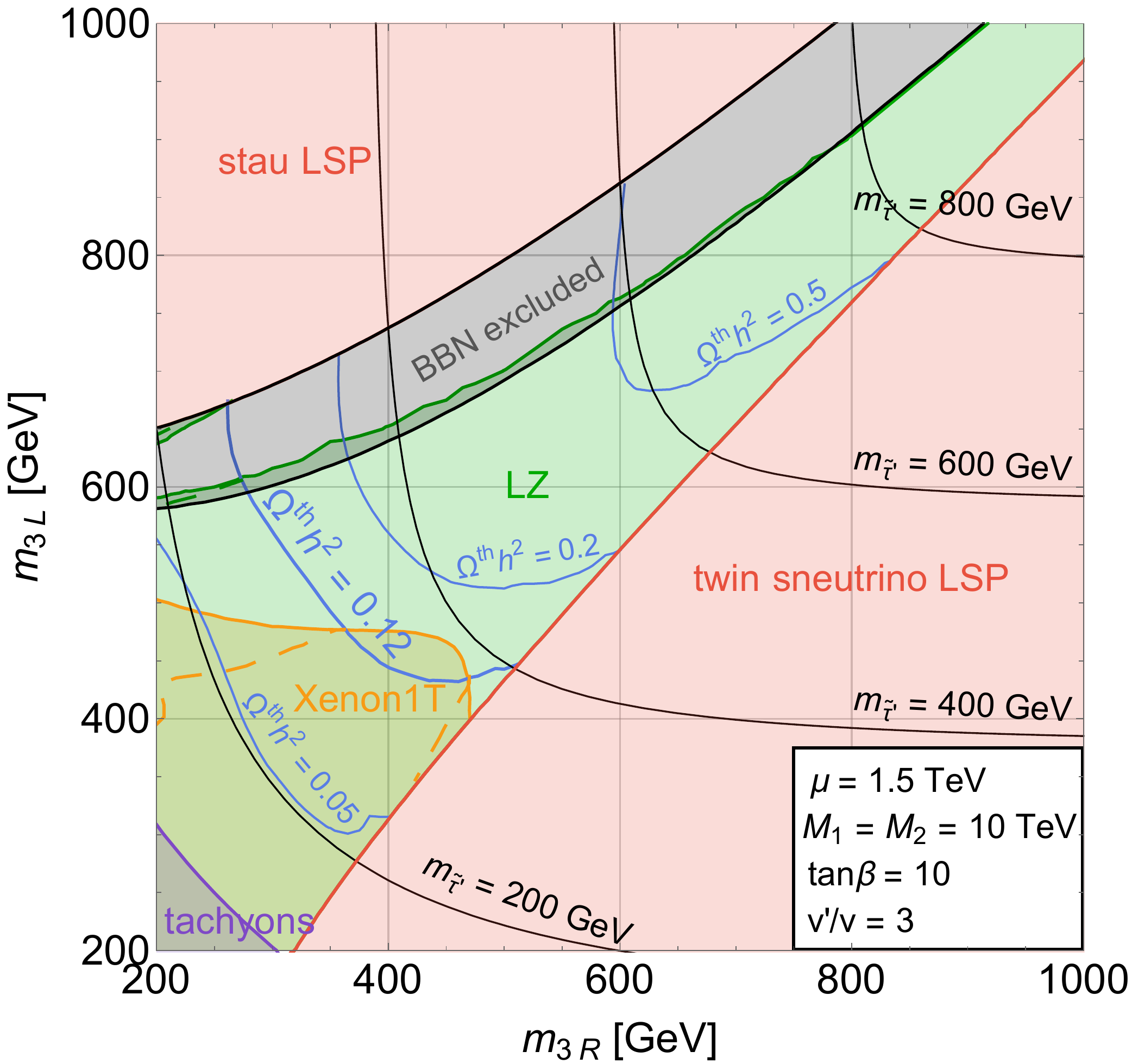}
 \caption{Contours of $\Omega^{\rm th} h^2$
 (blue lines) and the lightest twin stau mass (black line) in the plane of the right and left soft stau masses $(m_{3R},\,m_{3L}$, respectively) for $\mu=1.5$ TeV, $\tan\beta=10$, $v'/v = 3$ and the bino and wino mass parameters $M_1=M_2=10$ TeV. In the purple region there are tachyons in the spectrum and in the red regions $\tilde{\tau}'$ is not the LSP. The yellow (green) region denotes the area excluded (probed) by Xenon1T (LZ) assuming that $\Omega h^2=0.12$ in every point of the plane. The dashed curves instead assume the scaling of the bound with the thermal abundance if $\Omega^{\rm th} h^2<0.12$.
}
\label{fig:mLmR}
\end{center}
\end{figure} 
Fig.~\ref{fig:mLmR} shows the contours of $\Omega^{\rm th} h^2=0.12$ (blue) and the twin stau masses (black) in the twin stau soft masses parameter space. The correct thermal relic density is achieved for twin stau masses between about 300 and 400 GeV. In particular, in the region where the twin stau is mostly right-handed, coannihilation with the MSSM stau plays an important role in decreasing the effective cross section. As a consequence the thermal density is larger and $\Omega^{\rm th} h^2=0.12$ is obtained for smaller values of the SM mass. It is remarkable that the whole region preferred by the thermal abundance of the twin stau is consistent with astrophysical constraints on twin stau self-interactions which exclude masses below about 200~GeV.

Part of the parameter space is excluded simply due to the mass spectrum. The purple region leads to tachyons. In the red areas the twin stau is not the LSP; in the lower right region the twin sneutrino becomes the LSP, while in the upper left region the stau is lighter than the twin stau.

The yellow region is excluded by Xenon1T \cite{XENON:2018voc}, while the green region shows the potential reach of LUX-ZEPLIN (LZ) \cite{LUX-ZEPLIN:2018poe}. The dashed yellow (green) curve shows the direct detection bound from Xenon1T (potential reach of LZ) assuming thermal abundance. The yellow (green) solid curve, on the other hand, assumes that $\Omega h^2=0.12$ in the whole parameter space. This includes the case in which DM is produced non-thermally and the case in which the thermal abundance is too large but is reduced to the observed one because of entropy production. This plot shows that, in the scenario in which the thermal abundance of the LSP explains all of the dark matter, most of the parameter space is allowed by Xenon1T but most of it could be probed by LZ. In the thermal dark matter scenario with under-abundant LSP, which happens for twin stau masses smaller than $\sim300$ GeV, Xenon1T excludes significant portions of parameter space while LZ could probe most of it. Note, however, that LZ will not be sensitive enough to probe some region of parameter space corresponding to the LSP dominated by the right-handed twin stau. This region surrounds a blind spot in direct detection corresponding to a vanishing twin stau-Higgs coupling due to cancellation between the first and the last term in the squared bracket in eq.~\eqref{eq:hstau_coup}. However, in this region the decay $\tilde{\tau}\to \tilde{\tau}^{'\dag}\tau \tau'$ is kinematically closed and the stau lifetime is generically too long to be consistent with the BBN.%
\footnote{One may in principle make the stau short-lived enough to avoid the BBN constraints by $\mathcal{O}(1)$ mixing between the stau and the selectron or smuon. Using the computation in~\cite{Hisano:1995nq}, we find that the resultant flavor-changing decay rate of tau is below experimental upper bounds~\cite{BaBar:2009hkt}.}

We should note, however, that for the parameters chosen in Fig.~\ref{fig:mLmR}, in the absence of bino mass mixing, the decay length would be much above $1$ m because of the large $\mu$ and $M_1$. As a consequence, the stau would decay after the freeze-out, which can enhance the relic density. When the MSSM and twin staus are nearly degenerate with each other the contours of $\Omega^{\rm th} h^2$ remain almost unchanged while when their fractional mass difference is more than $1/x_f$, the enhancement factor is about 2 and requires a smaller stau mass. Furthermore, $m_{\tilde{\tau}} < 430$ GeV is excluded by the LHC searches for charged long-lived particles, as explained in section~\ref{sec:life}. Therefore, thermal dark matter case would be excluded. In order to avoid these conclusions it is enough to introduce the bino-twin bino mixing parameter $\epsilon_M\sim\mathcal{O}(10^{-1})$, see~eq.~\eqref{eq:Gamma_binomix}, since a typical mass splitting between twin stau and stau is $\mathcal{O}(10)$~GeV.

\begin{figure}
\begin{center}
 \includegraphics[width=0.495\textwidth]{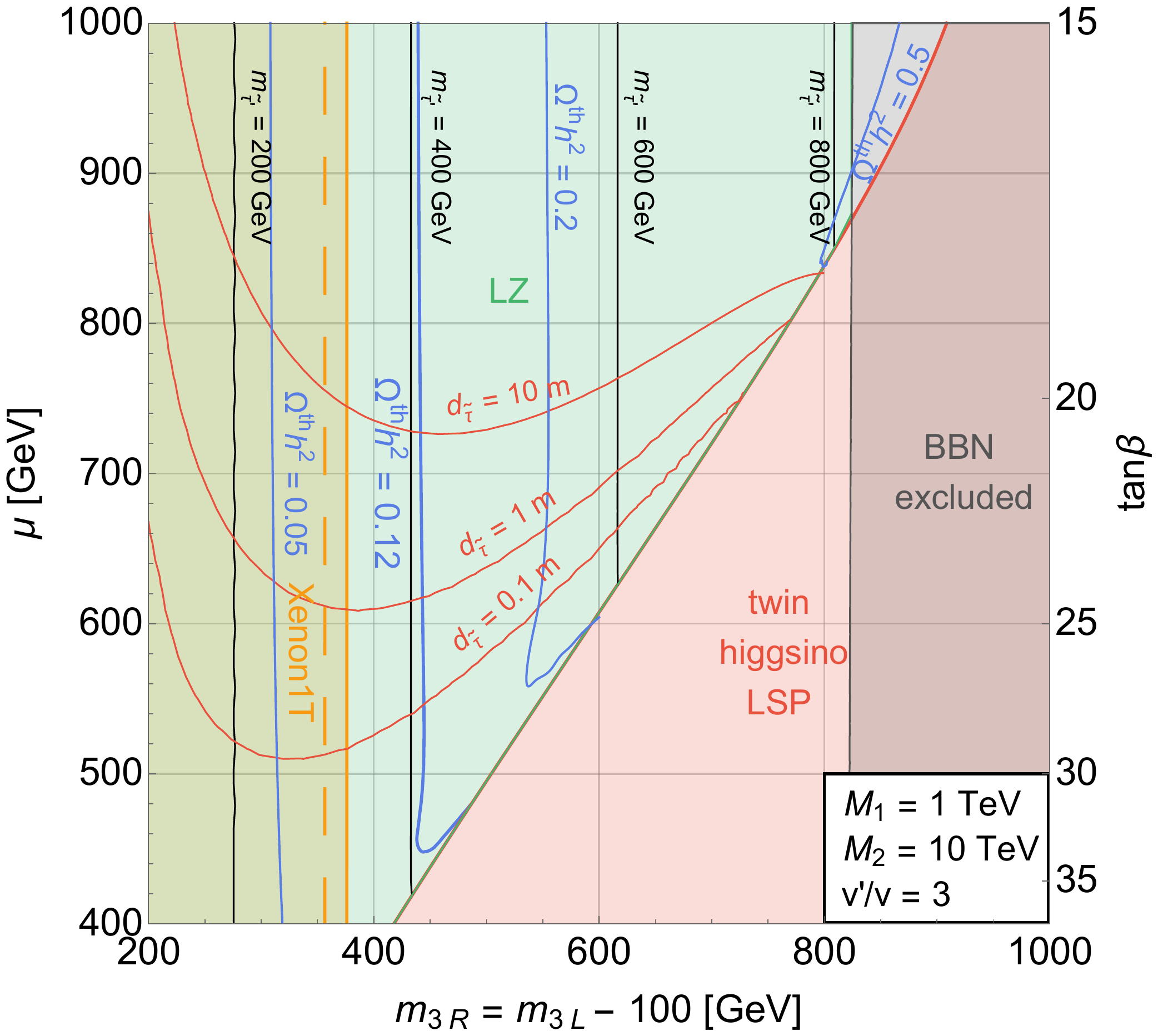}
 \includegraphics[width=0.46\textwidth]{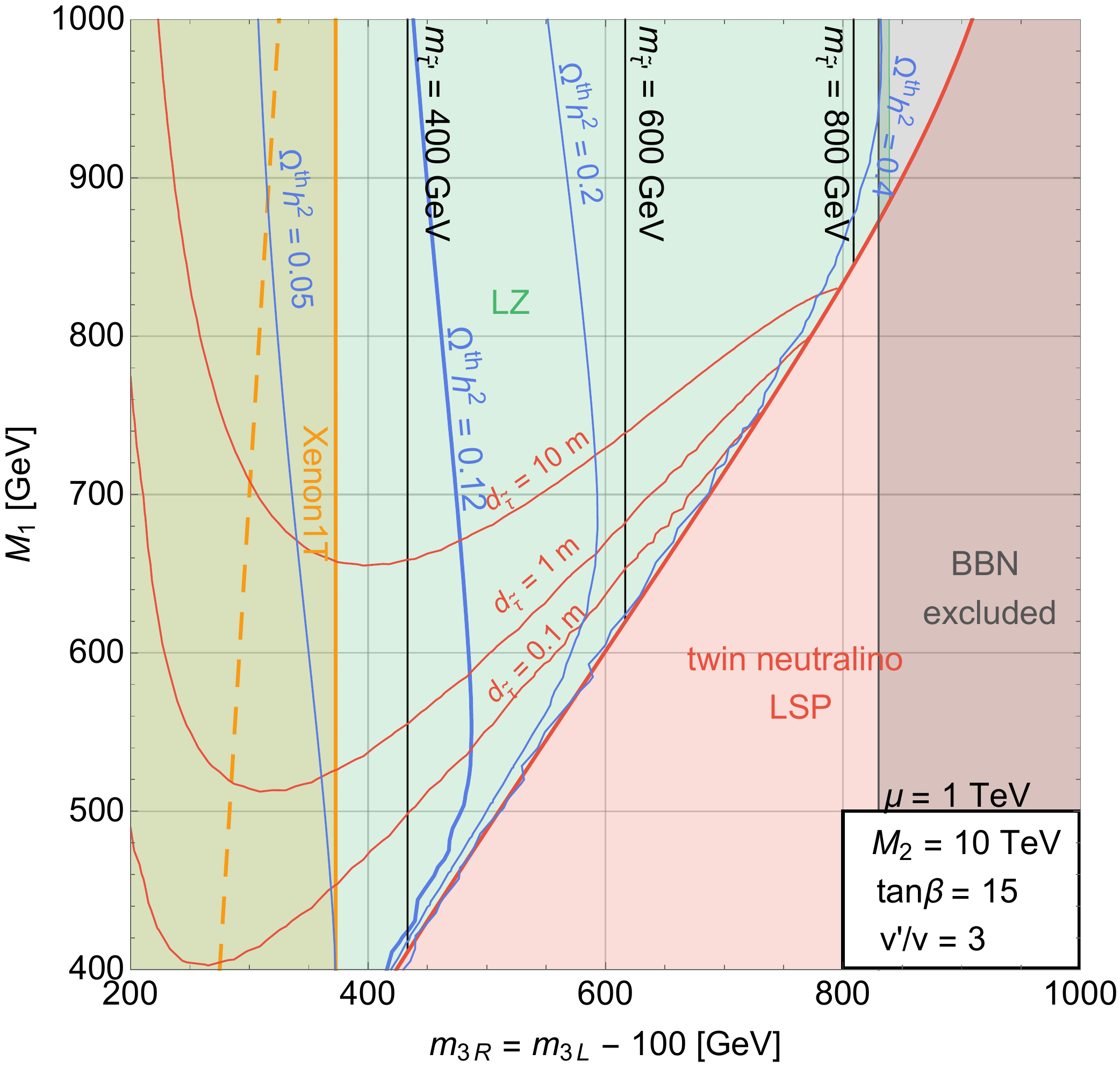}
 \caption{Contours of $\Omega^{\rm th} h^2$
 (blue lines) and the twin stau LSP mass (black line) in the plane ($m_{3R}, \mu$) for $\tan\beta=15\,\,\mathrm{TeV}/\mu$ (right axis) and $M_1=1$~TeV (left panel) and in the plane ($m_{3R}, M_1$) for $\mu=1$~TeV and $\tan\beta=15$ (right panel). The remaining parameters are set to $m_{3L} = m_{3R} + 100$ GeV, $v'/v = 3$, and $M_2=10$~TeV. In the red region the twin Higgsino (bino) is the LSP in the left (right) panel. The yellow and green regions are defined in the same way as in Fig.~\ref{fig:mLmR}. The red contours show the stau decay length $d_{\tilde{\tau}}$ assuming Higgsino mixing $\epsilon_H = 20$ GeV $(v'/v)/3$, which is typical in D-term Twin Higgs models. The mixing may be larger by an $\mathcal{O}(1)$ factor.
}
\label{fig:mu_mRmL100}
\end{center}
\end{figure}

In Fig.~\ref{fig:mu_mRmL100}, we show contours of $\Omega^{\rm th} h^2=0.12$ (blue) and $m_{\tstau}$ (black) in the $m_{3R}$ vs $\mu$ (left) and $m_{3R}$ vs $M_1$ (right) plane. We fix $M_2=10$ TeV, $v'/v=3$ and $m_{3L} = m_{3R} + 100$ GeV. In the left panel, in order to keep the stau mixing parameter fixed as $\mu$ varies, we take $\tan\beta= 15/\mu$ TeV, while $M_1=1$ TeV. The left panel shows that decreasing $\mu$ has no impact on the relic abundance of the twin stau unless the twin Higgsino is almost degenerate with the LSP, for which co-annihilation with the twin Higgsino becomes effective and the twin stau mass reproducing the observed dark matter abundance can be as large as about 450 GeV. For smaller $\mu$, the twin Higgsino becomes the LSP. The red curves show the contours of the decay length $d_{\tilde{\tau}} = 0.1$, $1$, and $10$ m via the Higgsino mixing discussed in Sec.~\ref{sec:life}. A large portion of the parameter space with $\mu\lesssim600$~GeV has $d_{\tilde{\tau}}$ below $1$ m. In such a region bino mass mixing is not necessary. We also note that the magnitude of the Higgsino mixing can be larger than our benchmark value by an $\mathcal{O}(1)$ factor, which can decrease the decay length by an $\mathcal{O}(10)$ factor and further expands the viable parameter space without resorting to bino mass mixing. For larger $\mu$ one has to still rely on bino mass mixing, although $\epsilon_M$ can be somewhat smaller, down to about $10^{-2}$, due to smaller $M_1$ than in Fig.~\ref{fig:mLmR}. 

In the right panel of Fig.~\ref{fig:mu_mRmL100} we show the dependence on $M_1$ for $\mu = 1$ TeV. The presence of a light bino enhances the twin stau annihilation cross-section into twin taus. As a result, for a sub-TeV bino $\Omega^{\rm th} h^2=0.12$ is obtained for a larger twin stau mass up to almost 500~GeV. In order to have the stau decay length below 1~m without introducing bino mixing, $M_1\lesssim600$~GeV is preferred for $\mu=1$~TeV. LZ will probe both the thermal and non-thermal scenario. In the non-thermal scenario, the region where the twin stau mass is above about 800~GeV will not be tested by LZ, but it is generically in conflict with the BBN. To suppress the stau decay length without introducing bino mass mixing, the bino mass is preferred to be below 800~GeV. 

\begin{figure}[t]
\begin{center}
 \includegraphics[width=0.49\textwidth]{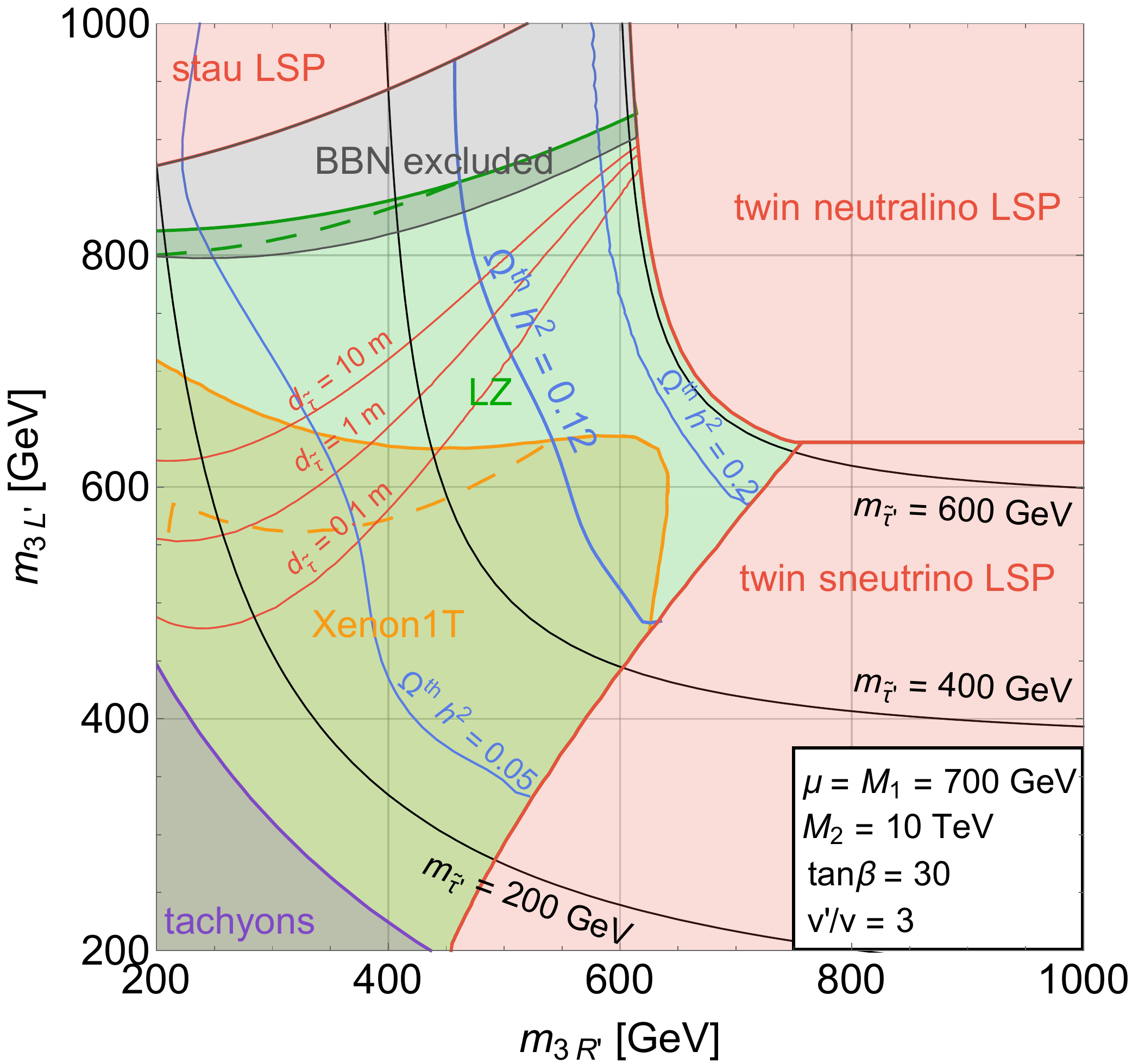}
  \includegraphics[width=0.49\textwidth]{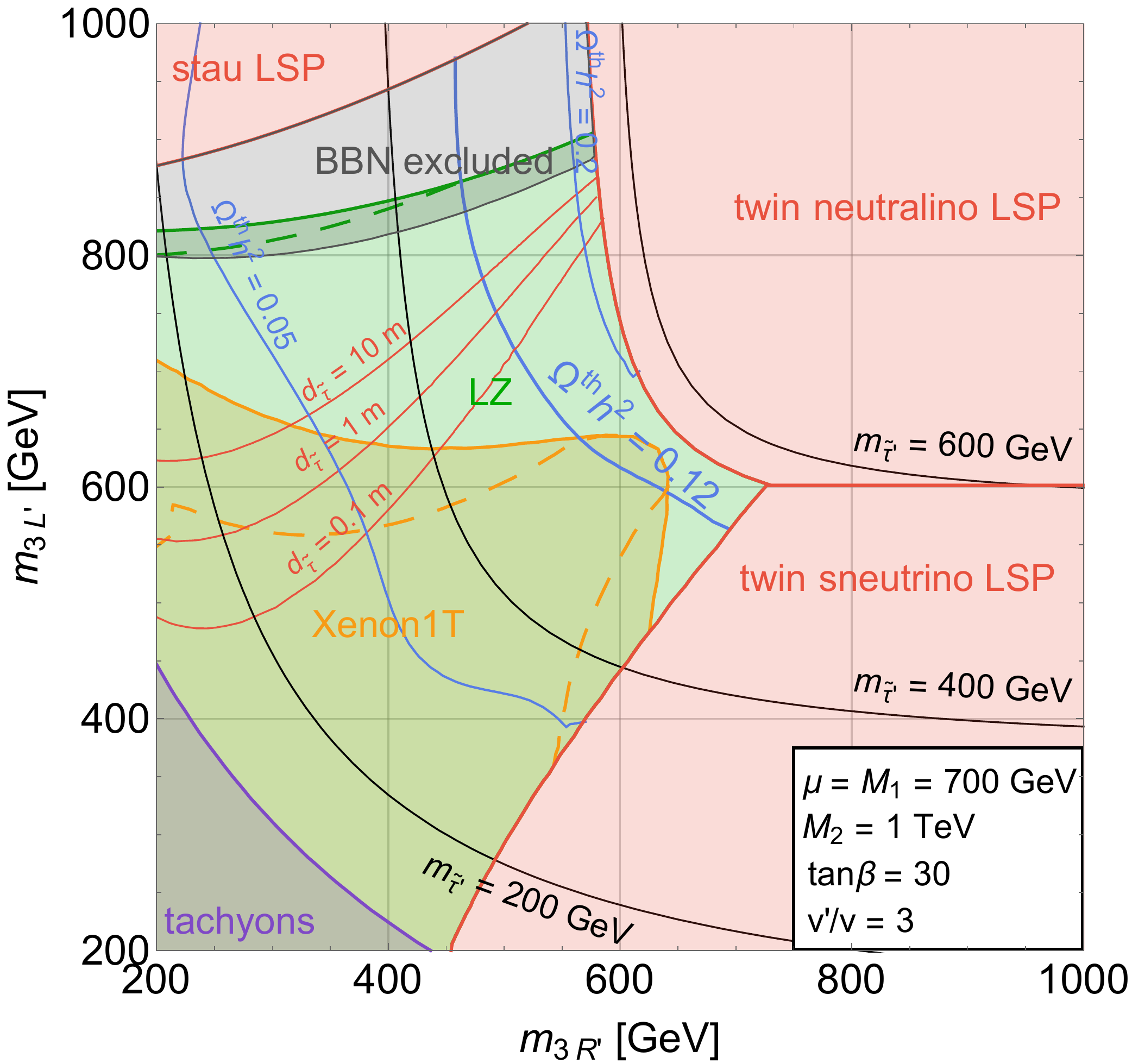}
 \caption{The same as in Fig.~\ref{fig:mLmR} but for $\mu=M_1=700$~GeV and $\tan\beta=30$. The left (right) panel has $M_2=10\,(1)$ TeV. The contours of the stau decay length assuming the Higgsino mass mixing are also shown.
}
\label{fig:mLmR_mu700}
\end{center}
\end{figure} 
In Fig.~\ref{fig:mLmR_mu700} we show how the stau parameter space opens up for $\mu=M_1=700$~GeV assuming $M_2=10~(1)$~TeV in the left (right) panel. In order to have large enough left-right stau mixing, we compensate the smaller value of $\mu$ by a large $\tan\beta=30$. A large portion of the parameter space has a stau decay length below 1~m and is allowed by Xenon1T. Moreover, in the region with the stau decay length above 1~m the mass of the twin stau playing the role of thermal DM is above 430 GeV, even after considering the enhancement of the relic abundance due to stau decays after the freeze-out. As a consequence, this region is allowed by the LHC constraints on long-lived charged particles. However, this requires also taking $\mu$ and $M_2$ above about 1.2 and 1.3~TeV respectively (increasing $\mu$ does not affect the relic abundance of the twin stau, and doing so for $M_2$ also does not in the parameter region with a stau decay length above 1~m), otherwise the production rate of long-lived staus from Higgsino and wino decays would exceed the rate allowed by the LHC searches. 

From the plots in Fig.~\ref{fig:mLmR_mu700} we can also see how a light wino impacts the relic abundance of the twin stau LSP. As expected, the contour of $\Omega^{\rm th} h^2=0.12$ does not change with a lighter wino as long as the LSP is dominated by the right-handed twin stau. However, in the vicinity of the twin sneutrino LSP region the left-handed stau component is non-negligible and the 1~TeV wino increases the twin stau mass corresponding to $\Omega^{\rm th} h^2=0.12$ up to about 550~GeV, compared to about 450~GeV in the case of decoupled wino. This also helps to avoid the Xenon1T constraint in that region and only a small part of the $\Omega^{\rm th} h^2=0.12$ contour is excluded.

\begin{figure}
\begin{center}
 \includegraphics[width=0.48\textwidth]{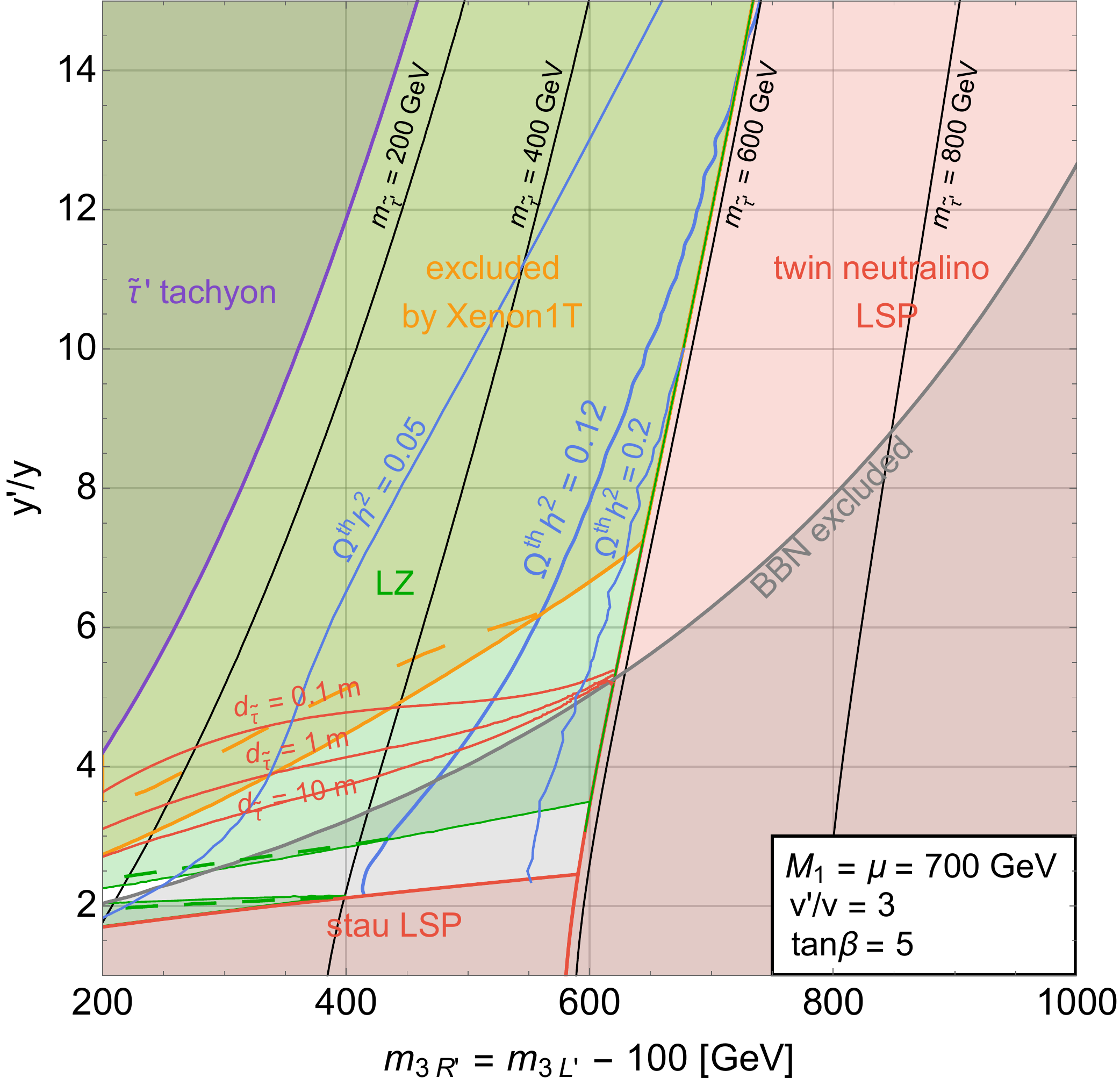}
 \caption{Contours of $\Omega^{\rm th} h^2$
 (blue lines) and the twin stau LSP mass (black line) in the plane ($m_{3R}, y_\tau'/y_\tau$). The remaining parameters are set to $m_{3L} = m_{3R}+100$ GeV, $\tan\beta=3$, $\mu=M_1=700$~GeV, $M_2=10$~TeV, and $v'/v = 3$.  The color coding of various regions is the same as in Fig.~\ref{fig:mLmR}. }
\label{fig:mu_mR_Atau_ytau}
\end{center}
\end{figure}  

We have been assuming so far strict $\mathbb{Z}_2$ symmetry in the Yukawa couplings. Let us now discuss what happens if $\mathbb{Z}_2$ breaking in the tau Yukawa coupling is present. This breaking does not spoil the TH solution to the hierarchy but can help to solve the dark radiation problem of TH models ~\cite{Barbieri:2016zxn,Barbieri:2017opf}. A larger twin tau Yukawa coupling increases the left-right mixing of twin staus. Moreover, the larger mixing increases the mass splitting between the twin stau LSP and stau, which suppresses the stau lifetime. One consequence is that smaller values of $\tan\beta$ are now viable. The impact of larger $y_{\tau'}$ is shown in Fig.~\ref{fig:mu_mR_Atau_ytau} for $\tan\beta=5$. For $y_{\tau'}/y_{\tau}$ above about two, the twin stau can be the LSP. However, in order to have the correct thermal relic abundance and evade the BBN bounds, the ratio should be above four. Furthermore, the stau decay length can be sufficiently small for most of the parameter space with thermal DM without introducing bino mass mixing.

An important feature of the scenario with an enhanced twin tau Yukawa coupling is that a larger twin stau mass leads to $\Omega^{\rm th} h^2\approx0.12$, consequence of  the increased annihilation cross-section into SM-like Higgses and twin tops. The twin stau mass may be up to about 600 GeV, above which the twin neutralino becomes the LSP. However, there is an upper bound on $y_{\tau'}/y_{\tau}$ (of about six in the thermal scenario) set by Xenon1T constraints. This is because the direct-detection cross-section is enhanced by the larger twin tau Yukawa coupling. Nevertheless, even after taking into account the Xenon1T constraint the twin stau thermal DM can exceed 500 GeV. The whole thermal scenario will be probed by LZ also in this case.

The range of allowed $y_{\tau'}/y_{\tau}$ depends on $\tan\beta$. For smaller (larger) values of $\tan\beta$ the lower bound on $y_{\tau'}/y_{\tau}$ gets stronger (weaker) while the upper bound from Xenon1T gets weaker (stronger). The stau decay length in the parameter space with thermal twin stau DM varies by orders of magnitude even without introducing bino-twin bino mixing. Some of the parameter space will be probed by disappearing track searches at the LHC but promptly decaying staus are possible for large enough $y_{\tau'}/y_{\tau}$.

\section{Dark radiation and dark matter abundance}
\label{sec:DR}

It is well known that Twin Higgs models in their simplest realizations suffer from too large amount of dark radiation. In particular, in Mirror Twin Higgs models $\Delta N_{\rm eff}\approx5.6$~\cite{Chacko:2016hvu}. This is order of magnitude above the upper bound from Planck of about $0.3$ at 95\%~C.L.~\cite{Planck:2018vyg}. Possible solutions to this problem are:
\begin{itemize}
    \item late entropy production which dilutes dark radiation~\cite{Chacko:2016hvu,Craig:2016lyx},
    \item $\mathbb{Z}_2$ breaking in the Yukawa couplings of light fermions~\cite{Barbieri:2016zxn,Barbieri:2017opf}, which reduces the entropy of the mirror sector by the Boltzman suppression,
    \item Fraternal Twin Higgs~\cite{Craig:2015pha} where the first and second generation mirror fermions are removed (the mirror photon, however, may be still in the spectrum).
\end{itemize}
In this section, we discuss the compatibility of the parameter space of the twin stau LSP DM with these solutions.

In Figs.~\ref{fig:mLmR}, \ref{fig:mu_mRmL100}, and \ref{fig:mLmR_mu700}, we assume $\mathbb{Z}_2$ symmetric yukawa, so only late time entropy production and Fraternal Twin Higgs are applicable. In the Fraternal Twin Higgs solution, the thermal abundance is maintained, so the prediction shown by the blue lines is applicable.

With entropy production, the twin stau abundance is affected. If the production of the twin stau is negligible during entropy  injection, which is guaranteed when the particle responsible for it is lighter than the twin stau, the twin stau abundance is simply diluted. As a consequence, the abundance consistent with the observed value would require a larger mass of the twin stau DM. The thermal abundance is roughly proportional to $m_{\rm LSP}^2$. Assuming that $\Delta N_{\rm eff}$ is diluted by a factor $D^{4/3}$, the DM abundance is reduced by a factor $D$ and we can link $\Delta N_{\rm eff}$ with the twin stau LSP mass that gives the observed DM abundance after taking into account entropy production:
\begin{equation}
\label{eq:Neff}
    \Delta N_{\rm eff}\approx0.3\left(\frac{m_{\tilde{\tau'}_1}}{3m_{\tilde{\tau'}_1}^{\rm th}}\right)^{-8/3} \,,
\end{equation}
where $m_{\tilde{\tau'}_1}^{\rm th}$ is the twin stau mass that gives thermal abundance equal to the observed value. Here we consider the limit where the heating of the twin sector is negligible during the entropy production. If not, $\Delta N_{\rm eff}$ is even larger. We see that in order to satisfy the Planck constraint on $\Delta N_{\rm eff}$, the twin stau DM mass would be around 1 TeV in the case of decoupled bino and even heavier for lighter bino. Interestingly, $\Delta N_{\rm eff}$ will be tested with much better precision and the bound on $\Delta N_{\rm eff}$ is expected to be set by CMB-S4 experiments~\cite{CMB-S4:2016ple} around $0.06$ at 95\%~C.L.~\cite{Baumann:2017gkg}. If CMB-S4 does not find signals of dark radiation this would set the lower bound on the twin stau mass of almost 2 TeV. This would result in an even heavier Higgsino, leading to fine-tuning of the EW scale worse than 1\%.

If the particle responsible for entropy production is heavier than the twin stau, twin staus may be produced during entropy production. The stau mass may be smaller than that predicted from Eq.~(\ref{eq:Neff}). It is even possible that the stau mass is smaller than the prediction of the thermal DM scenario shown by blue lines.

In Fig.~\ref{fig:mu_mR_Atau_ytau}, since we assume $\mathbb{Z}_2$ breaking yukawas, all of the solutions are applicable, and the case with full mirror sector and without entropy production would be the minimal one. Then the thermal abundance is maintained and the prediction shown by the blue lines is applicable. One can  in  addition introduce entropy production, for which case the stau mass may be different from that prediction. Because of the $\mathbb{Z}_2$ breaking in the yukawa, Eq.~(\ref{eq:Neff}) is no longer applicable.

\section{Conclusions}
\label{sec:conclusion}
We have investigated the possibility that DM in the Universe is a charged particle in supersymmetric Twin Higgs models. We found that the twin stau is a perfectly viable candidate for such a DM particle. In order to ensure that the twin stau is the lightest supersymmetric particle and hence stable, it must be a mixture of the right and left components. We found that the thermal relic abundance of the twin stau LSP matching the observed DM density implies that the twin stau mass is generically in the range of 300-400 GeV, but it can even exceed 500 GeV if some of (twin) neutralinos are not much heavier than the LSP. Intriguingly, the preferred range of the twin stau mass is above a lower bound of about 200~GeV from twin stau self-interactions arising from the non-zero twin electromagnetic charge of the twin stau.

Signatures of twin stau DM are very different from standard supersymmetric DM candidates. The scattering cross-section off nuclei for the twin stau is naturally suppressed because the twin stau interacts with nuclei only via the mixing of the twin Higgs with SM-like Higgs. As a result, the latest constraints from direct detection experiments are satisfied in most of the parameter space in contrast to standard supersymmetric scenarios in which fine-tuning of the parameters is necessary (coannihilation~\cite{Ellis:1998kh,Ellis:1999mm} and blind-spots~\cite{Cheung:2012qy,Huang:2014xua,Badziak:2015exr} require fine-tuning in the parameters relevant for the DM phenomenology, while a pure Higgsino or wino requires fine-tuning of the EW scale.). Nevertheless, the Higgs-twin Higgs mixing cannot be arbitrarily small in natural Twin Higgs models so a signal of twin stau dark matter is expected at the LZ experiment. The main distinctive feature of this scenario are self-interactions of the LSP which is not present in standard supersymmetric models and future astrophysical data may differentiate between these classes of DM. 

In this scenario the mass splitting between the twin stau LSP and the MSSM stau is generically small, below few tens of GeV. This may lead to long lifetime of the stau which, in turn, may have important implications for cosmology and collider searches for staus. The lifetime of the stau is model-dependent and can be short enough to avoid all cosmological and collider constraints. Nevertheless, in some of the parameter space one may expect signals of long-lived staus at the LHC, e.g., in disappearing track searches.

\section*{Acknowledgments}

This work was partially supported by the National Science Centre, Poland, under research grant no. 2020/38/E/ST2/00243 (M.B. and M.L.), 
by the INFN Iniziativa Specifica TAsP Theoretical Astroparticle Physics (G.G.d.C.)  
and by Friends of the Institute for Advanced Study (K.H.). 
G.G.d.C. is also supported by the Frascati National Laboratories (LNF) through a Cabibbo Fellowship, call 2019.

\appendix

\section{Higgsino mixing}
\label{sec:App_higgsinomix}

In this section, we derive Higgsino-twin Higgsino mixing in a D-term Twin Higgs model. We consider a $U(1)_X$ gauge symmetry with the corresponding gauge coupling $g_X$ and charges
\begin{align}
H_u(1/2),~H_u'(1/2),~H_d(-1/2),~H_d'(-1/2).
\end{align}
The gauge symmetry is broken by VEVs of $S$ and $\bar{S}$ with charges $1/2$ and $-1/2$, respectively. The VEVs may be induced by a superpotential
\begin{align}
    W = \lambda Y(S\bar{S}-M^2),
\end{align}
where $Y$ is a chiral multiplet. We assume the same soft masses of $S$ and $\bar{S}$ so that $\vev{S}=\vev{\bar{S}} \equiv v_S ( = \mathcal{O}(M))$ and the D-term potential of $U(1)_X$ does not give large Higgs masses. The mass of the $X$ gauge boson is $g_X v_S$. $Y$ obtains a tadpole term $\sim \lambda m_{\rm soft} v_S^2 Y$ via supersymmetry breaking. It also obtains a positive mass squared $=2 \lambda^2 v_{S}^2 |Y|^2$. From these two terms, $\lambda v_Y \sim m_{\rm soft}$.

The mass terms of relevant fermions are
\begin{align}
    {\cal L} = - \lambda v_S \tilde{Y}(\tilde{S} + \tilde{\bar{S}} ) - \lambda v_Y \tilde{S}\tilde{\bar{S}} - \frac{g_X}{\sqrt{2}}\tilde{X} \left(v_u\tilde{H}_u - v_d \tilde{H}_d + v_S\tilde{S} - v_S \tilde{\bar{S}} \right) - \frac{1}{2}m_{\tilde{X}} \tilde{X}\tilde{X} + {\rm h.c.},
\end{align}
where $m_{\tilde{X}}$ is the SUSY breaking Majorana mass of $\tilde{X}$. After integrating out $\tilde{X}$, $\tilde{S}$, and $\tilde{\bar{S}}$, the mixing between the MSSM and twin Higgsinos is given by
\begin{align}
    {\cal L} = \frac{\lambda v_Yg_X^2 v_u v_u'}{2 (g_X^2 v_S^2 + \lambda v_Y m_{\tilde{X}})} (\tilde{H}_u - \frac{1}{\tan\beta} \tilde{H}_d)(\tilde{H}_u' - \frac{1}{\tan\beta} \tilde{H}_d').
\end{align}
To have a non-decoupling D-term potential of the SM and twin Higgses from $U(1)_X$, we need $m_{\rm soft} \sim m_X = g_X v_S$.
The parameter $\epsilon_{\tilde{H}}$ in Eq.~(\ref{eq:higgsino_mixing}) is then estimated as
\begin{align}
    \epsilon_{\tilde{H}}  \sim \frac{m_{\rm soft} v_u v_u'}{2v_S^2 } \sim \frac{g_X v v'}{2v_S},
\end{align}
where we used $\lambda v_Y \sim m_{\rm soft}$ and neglected $m_{\tilde{X}}$. The electroweak precision measurements requires that $v_S > 4$ TeV~\cite{Badziak:2017kjk}. In our numerical calculations of the stau decay length we take, for concreteness, $g_X=2$, $v_S=5$~TeV and $\lambda v_Y = m_{\rm soft}=m_X=10$~TeV which leads to $\epsilon_{\tilde{H}}\approx 20~{\rm GeV} (v'/v)/3$ but we should keep in mind that $\epsilon_{\tilde{H}}$ can vary by $\mathcal{O}(1)$ factor depending on the soft masses of the $U(1)_X$ symmetry breaking sector.

\section{Generation of bino-twin bino mixing}
\label{sec:App_binomix}

In this appendix, we analyze a setup with a significant bino-twin bino mass mixing but with a negligible kinetic mixing.
We introduce a gauge singlet chiral multiple $Z$ and the following interaction,
\begin{align}
\int {\rm d}\theta^2 \frac{Z}{M_*} W W'.
\end{align}
This interaction preserves the following approximate $Z_2\times Z_2$ symmetry,
\begin{align}
    Z \rightarrow - Z,&~~V\rightarrow -V \\
     Z \rightarrow - Z,&~~V'\rightarrow -V', \\
\end{align}
where $V$ is a hypercharge gauge multiplet.

We may give a non-zero $F$ term to $Z$ while fixing it near the origin via, for example, the O’Raifeartaigh model~\cite{ORaifeartaigh:1975nky},
\begin{align}
    W = \Lambda^2 Z + m X Y + \frac{k}{2} Z X^2,
\end{align}
where $X$ and $Y$ are chiral multiplets. The $Z_2$ symmetry is promoted to $Z_4$ symmetry with $Z\rightarrow -Z$, $X\rightarrow i X$ and $Y \rightarrow -i Y$.
A dimensionful parameter $\Lambda$ is a spurion of $Z_4$ symmetry breaking. We may obtain it dynamically by embedding this into the IYIT dynamical supersymmetry breaking model~\cite{Izawa:1996pk,Intriligator:1996pu}, where $m\sim \Lambda$ is also generated dynamically.
$Z$ obtains an $F$ term of $\Lambda^2$ while is fixed at the origin by the quantum correction from $X$ and $Y$,
\begin{align}
    K \simeq - \frac{k^4}{16\pi^2 m^2} ZZ^\dag Z Z^\dag.
\end{align}
$Z$ obtains a mass term $\sim k^4 \Lambda^4/(16\pi^2 m^2)$ and a tadpole term $\sim \Lambda^2 m_{3/2} Z$, so obtains a non-zero VEV $\sim 16\pi^2 m^2 m_{3/2} / (k^4 \Lambda^2)$. For $k\sim 1$ and $m\sim \Lambda$, the resultant kinetic mixing is much smaller than the experimental upper bound.

Explicit breaking of the symmetry also induces a non-zero field value of $Z$.
Indeed, the action on the hypercharge gauge field is charge-conjugation symmetry, which is explicitly broken by the electroweak gauge interaction. Quantum corrections to $D$ terms of $V$ and $V'$ gives $Z m_{\rm soft}^4/(256\pi^4 M_*)$. There also exists a tree-level contribution from the Higgs VEVs $\sim Z v^2 v'^2/ M_*$. The VEV of $Z$ induced by these tadpole terms is sufficiently small.

\bibliographystyle{JHEP}
\bibliography{draft}

\end{document}